\documentstyle[12pt,aasms4]{article}
\newcommand{\onu}{\Omega_\nu}
\newcommand{\om}{\Omega_{\rm m}}
\newcommand{\ov}{\Omega_\Lambda}
\newcommand{\ob}{\Omega_{\rm b}}

\newcommand{\xinl}{\bar\xi_{\rm nl}}
\newcommand{\xil}{\bar\xi_{\rm l}}

\newcommand{\Dnl}{\Delta_{\rm nl}}
\newcommand{\Dl}{\Delta_{\rm l}}
\newcommand{\pd}{Peacock \& Dodds\ }
\def\go{\mathrel{\raise.3ex\hbox{$>$}\mkern-14mu
\lower0.6ex\hbox{$\sim$}}}
\def\lo{\mathrel{\raise.3ex\hbox{$<$}\mkern-14mu
\lower0.6ex\hbox{$\sim$}}} \def\onu{\Omega_\nu}

\begin{document}
\title{Analytical Approximation to the Nonlinear Power Spectrum of
Gravitational Clustering}
\author{Chung--Pei Ma} 
\affil{Department of Physics and Astronomy,
University of Pennsylvania, Philadelphia, PA~19104}
\authoremail{cpma@strad.physics.upenn.edu}

\begin{abstract}  
This paper presents an analytical formula that closely approximates
the fully nonlinear power spectrum of matter fluctuations for redshift
$z\approx 5$ to 0 over a wide range of cosmologically interesting flat
models with varying matter density $\Omega_m$ and neutrino fraction
$\onu$.  The functional form is motivated by analytical solutions in
asymptotic regimes, but in order to obtain accurate approximations,
the coefficients are calculated from fits to the nonlinear power
spectrum computed from numerical simulations of four cosmological
models.  The transformation from the linear to nonlinear power
spectrum depends on $\om$, $\onu$, and time.  A simple scaling rule is
introduced, which greatly simplifies the construction of the
functional form and allows the formula to depend directly on the rms
linear mass fluctuation $\sigma_8$ instead of on an effective spectral
index as in earlier work.
\end{abstract}


\section{Introduction}
The power spectrum of matter fluctuations $P(k)$ provides the most
fundamental statistical measure of gravitational clustering.  When the
amplitude of the density fluctuations is small, the power spectrum can
be calculated easily from the linear perturbation theory of
gravitational collapse.  In the nonlinear regime, however,
determination of the fully evolved power spectrum for a given
cosmological model requires numerical simulations.  Since nearly all
observable astronomical systems have experienced some nonlinear
collapse, it would provide much physical insight and practical
convenience to devise a general analytical approximation (based on
simulation results) for the nonlinear power spectrum for a wide range
of cosmologically interesting models.

By fitting to $N$-body results, Hamilton et al. (1991) studied
scale-free models with a power-law spectrum and presented a simple
analytical formula that relates the spatially averaged nonlinear and
linear two-point correlation function, $\xinl(r)$ and $\xil(r_0)$,
where $r$ is related to its pre-collapsed linear scale $r_0$ by
$r_0=r\,(1+\xinl)^{1/3}$.  This transformation then appeared to be
magically independent of the spectral index $n$ assumed in the model.
Further tests against numerical simulations, however, found
significant errors when the Hamilton et al. function was applied to
models with $n<-1$ (Jain, Mo, \& White 1995; Padmanabhan 1996).  Jain
et al. (1995) instead proposed $n$-dependent formulas to relate
$\bar\xi$ and $P(k)$ in the linear and nonlinear regimes in both
scale-free models and the standard CDM model.  For the more realistic
CDM model, for which the spectral index changes from the primordial
value $n\approx 1$ on large scales to nearly $-3$ on small scales,
they used an effective index given by $n_{\rm
eff}=d\ln\,P(k)/d\ln\,k|_{k_c}$, where $k_c$ is the scale at which the
rms mass fluctuation $\sigma$ is unity.  The index $n_{\rm eff}$
therefore reflects the slope of the power spectrum at the length scale
where nonlinearity becomes important.  Peacock and Dodds (1996)
extended this work to allow for a low $\om$ and a non-zero
cosmological constant.

No previous work has investigated in detail the subject of linear to
nonlinear mapping in cold+hot dark matter (C+HDM) models that assume
massive neutrinos are a component of the dark matter.  This is perhaps
because the physics in C+HDM models is generally more complicated than
in CDM or LCDM (CDM with a cosmological constant) models due to the
additional length scale associated with the free streaming of the
neutrinos (Ma 1996).  Nevertheless, massive neutrinos remain a prime
dark matter candidate, and the recent evidence for neutrino masses from 
the Super-Kamiokande experiment has made this possibility
particularly intriguing (Fukuda et al. 1998).  Although neither Jain
et al. or \pd has tested these models, one may surmise that their
formulas can be naturally extended to C+HDM models as long as the
spectral index in the formula is calculated from the C+HDM power
spectrum.  This unfortunately does not work.  Both fitting functions
underestimate the nonlinear density variance,\footnote{The notations
$\Delta$ and $P(k)$ here are the same as in Jain et al., and are
equivalent to $\Delta^2$ and $P(k)/(2\pi)^3$ in \pd.} $\Dnl=4\pi
k^3\,P_{\rm nl}(k)$, at $k\go 2\,h$ Mpc$^{-1}$ in C+HDM models, and
the errors in \pd, for example, reach $\sim 50$\% at $k\sim 10\,h$
Mpc$^{-1}$.  The linear to nonlinear transformation is therefore
regulated by more than simply $n_{\rm eff}$.  (Smith et al. (1998)
recently reported agreement between the Peacock-Dodds formula and
results from two C+HDM simulations.  Their simulation resolution of
$\sim 0.3\,h^{-1}$ Mpc, however, limited their test to only the mildly
nonlinear regime, and could not probe the nonlinear regime where the
large discrepancies reside.)

This {\it Letter} differs from previous work in two ways.  First, the
simple analytical formula presented here closely approximates the
fully nonlinear power spectrum of mass fluctuations at $z\lo 5$ in the
previously unexplored C+HDM models as well as LCDM models with varying
$\om$.  Numerical simulations of four COBE-normalized flat C+HDM and
flat LCDM models are performed to calibrate the coefficients in the
analytical formula.  Second, the formula introduced here depends
directly on $\sigma_8$ (the rms linear mass fluctuation on $8\,h^{-1}$
Mpc scale) instead of a spectral index as in previous work.  This is
achieved by recognizing a scaling rule (see \S~3), which also greatly
simplifies the construction of the analytical formula.  This work also
extends into the nonlinear regime a previous investigation of the
effects of neutrino free-streaming on the linear C+HDM power spectrum
(Ma 1996).

\section{Input Linear Power Spectrum}
For a wide range of CDM and LCDM models that assume neutrinos are
massless, a good approximation to the linear power spectrum is given
by
\begin{equation}
     P(k,a,\onu=0) = 
    { A\,k^n\,[D(a)/D_0]^2\,\left[{\ln(1+\alpha_1 q)/ \alpha_1 q}\right]^2 
	 \over [1+\alpha_2 q+(\alpha_3 q)^2+(\alpha_4 q)^3
	+(\alpha_5\,q)^4]^{1/2}} \,,
\label{bbks}
\end{equation} 
where $k$ is the wavenumber in units of Mpc$^{-1}$, $q=k/\Gamma\,h$,
$\Gamma$ is a shape parameter, and $\alpha_1=2.34, \alpha_2=3.89,
\alpha_3=16.1, \alpha_4=5.46$, and $\alpha_5=6.71$ (Bardeen et
al. 1986).  The shape parameter $\Gamma$ characterizes the dependence
on cosmological parameters and is well approximated by $\Gamma=\om
h/\exp[\ob(1+1/\om)]$ (Efstathiou et al. 1992; Sugiyama 1995; see also
Bunn \& White 1997).  The function $D(a)$ is the linear growth factor,
whose present value is $D_0=D(a=1)$, and it can be expressed as
$D(a)=a\,g$, where the relative growth factor $g$ is well approximated
by $g(\om(a),\ov(a)) =2.5\,\om(a) [ \om(a)^{4/7}-\ov(a)+\left(1+
\om(a)/2\right) \left(1+ \ov(a)/70 \right) ]^{-1}$ (Lahav et al. 1991;
Carroll et al. 1992).  In LCDM models, $g\approx 1$ until the universe
becomes $\Lambda$-dominated at $1+z\approx \om^{-1/3}$; the value of
$g$ then decreases with increasing $a$.  The normalization factor $A$
can be chosen by fixing the value of $\sigma_8$; if instead the COBE
normalization is desired, it is $A=\delta_H^2 (c/H_0)^{n+3}/(4\pi)$,
where (for flat models) $\delta_H=$ $1.94\times
10^{-5}\,\om^{-0.785-0.05\,\ln\,\om} \,\exp(-0.95 \tilde{n}-0.169
\tilde{n}^2)$ with $\tilde{n}=n-1$ (Bunn \& White 1997).

The linear power spectra for the C+HDM models require additional
treatment since the effect of massive neutrinos on the shape of the
power spectrum is both time and scale dependent.  It is found that by
introducing a second shape parameter, $\Gamma_\nu=a^{1/2}\onu h^2$, to
characterize the neutrino free-streaming distance, one can obtain a
good approximation to the linear power spectra (density averaged over
the cold and hot components) in flat C+HDM models
at $z\lo 5$ when neutrinos are adequately nonrelativistic (Ma 1996):
\begin{equation}
   P(k,a,\onu)=P(k,a,\onu=0)
	\left( { 1+d_1\,x^{d_4/2}+d_2\,x^{d_4} \over 1+d_3\,x_0^{d_4} }
	\right)^{\onu^{1.05}} \,,
\label{pave}
\end{equation}
where $x=k/\Gamma_\nu\,$, $x_0=x(a=1)\,$, $P(k,a,\onu=0)$ for the pure
CDM model is given by eq.~(\ref{bbks}), and $d_1=0.004321,
d_2=2.217\times 10^{-6}, d_3=11.63$, and $d_4=3.317$ for $k$ in
Mpc$^{-1}$.  (The scale factor in $\Gamma_\nu$ assumes COBE
normalization at $a=1$.  If other normalization is used, replace $a$
by $a\,\sigma_8/\sigma_8^{\rm cobe}$.)

\section{Nonlinear Power Spectrum}
Numerical simulations of structure formation in two flat C+HDM models
with neutrino fraction $\onu=0.1$ and 0.2 and two flat LCDM models
with matter density $\om=0.3$ and 0.5 are performed in order to obtain
the nonlinear power spectra of matter fluctuations.  All four
simulations are performed in a (100 Mpc)$^3$ comoving box.  The
gravitational forces are computed with a particle-particle
particle-mesh (P$^3$M) algorithm (Bertschinger \& Gelb 1991; Ma et
al. 1997) with a comoving Plummer force softening length of 50 kpc.
An identical set of random phases is used in the initial conditions
for all four runs.  The primordial power spectrum has an index
of $n=1$, with density fluctuations drawn from a random Gaussian
field.  A total of $128^3$ simulation particles are used to represent
the cold dark matter.  For the C+HDM models, $128^3$ and $10\times
128^3$ particles are used to represent the hot component in the
$\onu=0.1$ and 0.2 models, respectively.  Although the large particle
number is needed to finely sample the velocity phase space (Ma \&
Bertschinger 1994), tests performed for the $\onu=0.2$ model show that
the power spectrum itself is little affected when $128^3$ hot
particles are used.  Since structure forms too late in flat C+HDM
models with $\onu > 0.2$ (Ma et al. 1997 and references therein), only
the $\onu=0.1$ and 0.2 models are studied here.  Both models assume
$\ob=0.05$ and $h=0.5$, while $\Omega_{\rm cdm}=0.85$ and 0.75 for the
two models, respectively.  The two LCDM models chosen for the
simulations have $(\om,\ov,h)=(0.3,0.7,0.75)$ and $(0.5,0.5,0.7)$, and
$\ob=0$.  All four models are normalized to the 4-year COBE results
(Bennett et al. 1996; Gorski et al. 1996).

Figure~1 contrasts the linear (dotted) and nonlinear (solid) power
spectra at various redshifts for the four simulated models.  The
hierarchical nature of gravitational collapse in these models is
evident: the high-$k$ modes have become strongly nonlinear, whereas
the low-$k$ modes still follow the linear power spectrum.  The dashed
curves are from the analytical approximation described below.
Figure~2 illustrates the dependence of the linear to nonlinear
transformation on both time and cosmological parameters by plotting at
various redshifts the ratio of the nonlinear and linear density
variance, $\Dnl(k)/\Dl(k_0)$, against the linear $\Dl(k_0)$ (where
$\Delta=4\pi k^3 P$).  Note that $\Dnl$ and $\Dl$ are evaluated at
different wavenumbers, where $k_0=k\,(1+\Dnl)^{-1/3}$ corresponds to the
pre-collapsed scale of $k$, as introduced in Hamilton et al. (1991).
The dependence of $\Dnl(k)/\Dl(k_0)$ on $\om$, $\onu$, and time,
however, demonstrates that the universality seen by Hamilton et al. is
limited to the power-law models with $n>-1$ studied there.  The linear
to nonlinear mapping is clearly more complicated in more realistic
cosmological models (such as LCDM and C+HDM) whose spectral index
changes continuously with length scale.

Despite the apparent complication from the time dependence in
Figure~2, I find the scaling rule given below very helpful in
simplifying the construction of an analytical approximation.  As
Figure~3 shows, the time dependence is largely removed when $\Dnl/\Dl$
is plotted against the scaled quantity $\tilde{\Dl}$,
\begin{equation}
	\tilde{\Dl}={\Dl \over \sigma_8^\beta} \,, \quad \beta=0.7 +
	10\,\onu^2
\label{beta}
\end{equation}
instead of $\Dl$.  For example, $\beta=0.7$ for the CDM and LCDM
models, while $\beta=0.8$ and 1.1 for the $\onu=0.1$ and 0.2 C+HDM
models.  The deviations in the two LCDM models at late times ($a\go
0.7$) result from the retardation in the relative growth factor $g$ in
$\om<1$ models.  As shown in eq.~(\ref{master}) below, they can be
easily accounted for by including a factor of $g^3$ in the highly
nonlinear regime in the fitting formula.

The scaling behavior in eq.~(\ref{beta}) can be understood in terms of
the time dependence of the local slope of the linear power
spectrum near the scale where nonlinear effects become
important.  An example is the effective spectral index $n_{\rm
eff}+3=d\ln\,\Dl/d\ln\,k|_{k_c}$ considered by Jain et al. (1995),
where $k_c$ is the scale at which the rms linear mass fluctuation
$\sigma$ equals unity.
Figure~4 illustrates the dependence of $n_{\rm eff}$ on time and
cosmological parameters.  The value of $n_{\rm eff}$ generally
increases with time because as $\sigma$ grows, the wavenumber $k_c$ at
which $\sigma=1$ decreases, and the spectral index at $k_c$
(i.e. $n_{\rm eff}$) becomes larger since the slope of the power
spectrum for the models studied here always increases with decreasing
$k$.  In Figure~4, $n_{\rm eff}$ exhibits the fastest growth in the
$\onu=0.2$ C+HDM model at $\sigma_8\go 0.3$ because the neutrino free
streaming effect is more prominent in higher $\onu$ models, which acts
to suppress structure growth below the free streaming scale, causing
$\Dl$ to bend more at $0.1 < k < 1\,h$ Mpc$^{-1}$ (see the dotted
curves in Fig.~1).  At $\sigma_8\lo 0.3$ in the $\onu=0.2$ model, on
the other hand, $n_{\rm eff}$ stays nearly constant because it is
probing the nearly flat, $k>5\,h$ Mpc$^{-1}$ part of $\Dl$.  Despite
this interesting behavior, Figure~4 shows that for all models, the
dependence of $n_{\rm eff}+3$ on $\sigma_8$ is well approximated by a
power law at $\sigma_8\go 0.3$: $d\ln\,(n_{\rm eff}+3)/d\ln\,\sigma_8
\propto \beta$, where $\beta=0.7+10\,\onu^2$ as given in eq.~(3).
This therefore explains why replacing the factor $n_{\rm eff}+3$ in
the earlier work with $\sigma_8^\beta$ works well here.

The simple scaling behavior introduced by eq.~(\ref{beta}) allows one
to approximate the evolution of the nonlinear power spectrum directly
in terms of $\sigma_8$ instead of $n_{\rm eff}$.  Combining these
factors, I find that a close approximation for the nonlinear power
spectrum is given by
\begin{equation}
	{\Dnl(k)\over \Dl(k_0)} =
	G\left({\Dl \over g_0^{1.5}\,\sigma_8^\beta} \right) \,,\quad
	G(x)=[1+\ln(1+0.5\,x)]\,{1+0.02\,x^4 + c_1\,x^8/g^3 \over
	1+c_2\,x^{7.5}}\,,
\label{master}
\end{equation}
where $\beta$ is given by eq.~(3), $k_0=k\,(1+\Dnl)^{-1/3}$, and
$g_0=g(\om,\ov)$ and $g=g(\om(a),\ov(a))$ are, respectively, the
relative growth factor\footnote{ Even for C+HDM models, $g$ in eq.~(4)
is taken to be the familiar $g(\om(a),\ov(a))$ for $\onu=0$ models.
The true relative growth factor for C+HDM models is in fact given by
eq.~(2), but due to its complicated scale dependence, attempts thus
far to incorporate this factor directly in eq.~(4) have not led to
approximations with high accuracies (see Ma 1998).} at present day and
at $a$ discussed in \S~2.  The time dependence is in factors
$\sigma_8^\beta$ and $g$.  For CDM and LCDM, a good fit is given by
$c_1=1.08\times 10^{-4}$ and $c_2=2.10\times 10^{-5}$.  For C+HDM, a
good fit is given by $c_1=3.16\times 10^{-3}$ and $c_2=3.49\times
10^{-4}$ for $\onu=0.1$, and $c_1=6.96\times 10^{-3}$ and
$c_2=4.39\times 10^{-4}$ for $\onu=0.2$.  The dependence of $c_1$ and
$c_2$ on $\onu$ can in principle be cast in a functional form (see Ma
1998), but since the allowed range of $\onu$ is narrow, separate
coefficients are given here in order to obtain the highest possible
fitting accuracy.  The accuracy of eq.~(\ref{master}) is illustrated
in Figure~1 (dashed curves), where the rms error for each model ranges from 
3\% to 10\% for $k\lo 10\,h$ Mpc$^{-1}$ at all times except $z\go
4$, when the errors are about 15\%.

The functional form of $G(x)$ in eq.~(4) is chosen to give the
appropriate asymptotic behavior $\Dnl \rightarrow \Dl$ in the linear
regime ($x\ll 1$) and $\Dnl \propto \Dl^{3/2}$ in the stable
clustering regime ($x\gg 1$).  In the mildly nonlinear regime, $0.1<
\Dl < 1$, the pre-factor $[1+\ln(1+0.5\,x)]$ is introduced to
approximate the non-negligible positive slope of $\Dnl/\Dl$.  This
factor is needed because $\Dnl$ and $\Dl$ are evaluated at different
wavenumbers $k$ and $k_0$, where the pre-collapsed $k_0$ is always
smaller than $k$.  Due to the steep positive slope of $\Dl$ in this
region, $\Dl$ at $k_0$ is noticeably smaller than at $k$, and
$\Dnl(k)/\Dl(k_0)$ is thus significantly above unity.  Without the
logarithmic pre-factor in eq.~(\ref{master}) to account for this
elevation, the approximation to $\Dnl$ can be underestimated by up to
30\% at $0.1 <\Dl< 1$.

\section{Summary and Discussion}
This paper presents a single formula, eq.~(\ref{master}), that
accurately approximates the fully nonlinear power spectrum of matter
fluctuations for redshift $z\lo 5$ for flat CDM, LCDM, and C+HDM
models with varying matter density $\om$ and neutrino fraction $\onu$.
Eqs.~(1), (2) and (\ref{master}) together offer a complete description
of the shape and time evolution of the matter power spectrum in both
linear and nonlinear regime for a wide range of cosmologically
interesting models.  Figure~1 summarizes the analytical and simulation
results for four representative models.  Depending on the models and
epochs, the rms errors are between 3\% and 10\% for $k\lo 10\,h$
Mpc$^{-1}$ at $z\lo 4$ in eq.~(\ref{master}).  In comparison, the \pd
formula has a rms error between 6\% and 17\% for the two LCDM
simulations studied here, and the error reaches 50\% at $k\sim 10\,h$
Mpc$^{-1}$ for the two C+HDM models.

In contrast to the scale-free models studied in Hamilton et
al. (1991), the relation between the linear and nonlinear power
spectrum is far from universal.  The different panels of Figure~2
illustrate the dependence on cosmological parameters $\om$ and $\onu$.
Moreover, within a given model, the fact that the curves for different
times in Figure~2 do not overlap is important because it implies that
the ratio $\Dnl(k)/\Dl(k_0)$ depends not only on $\Dl(k_0)$ but also
on time, or equivalently, the overall amplitude of $\Dl(k_0)$.  This
amplitude (or time) dependence is present, albeit in a somewhat subtle
form, in earlier work.  The effective spectral index used by Jain et
al. (1995) is clearly time-varying as shown in Figure~4 of this paper.
The local spectral index $n=d\ln P/d\ln k|_{k_0}$ used by Peacock and
Dodds (1996) depends on time implicitly through the factor $\Dnl$ in
the relation $k_0=k\,(1+\Dnl)^{-1/3}$.  In comparison, in this paper I
have adopted the commonly used parameter $\sigma_8$ instead of a
spectral index to characterize this time dependence, and have shown
that the scaling behavior in eq.~(\ref{beta}) and Figure~3 absorbs
this dependence in C+HDM as well as LCDM models and results in the
simple formula, eq.~(4).

I thank the referee, Andrew Hamilton, for insightful comments that
have helped to improve the manuscript, and Robert Caldwell for
stimulating discussions.  Supercomputing time for the numerical
simulations is provided by the National Scalable Cluster Project at
the University of Pennsylvania and the National Center for
Supercomputing Applications.  A Penn Research Foundation Award is
acknowledged.


\clearpage
\begin{figure}
\epsfxsize=6.5truein 
\epsfbox{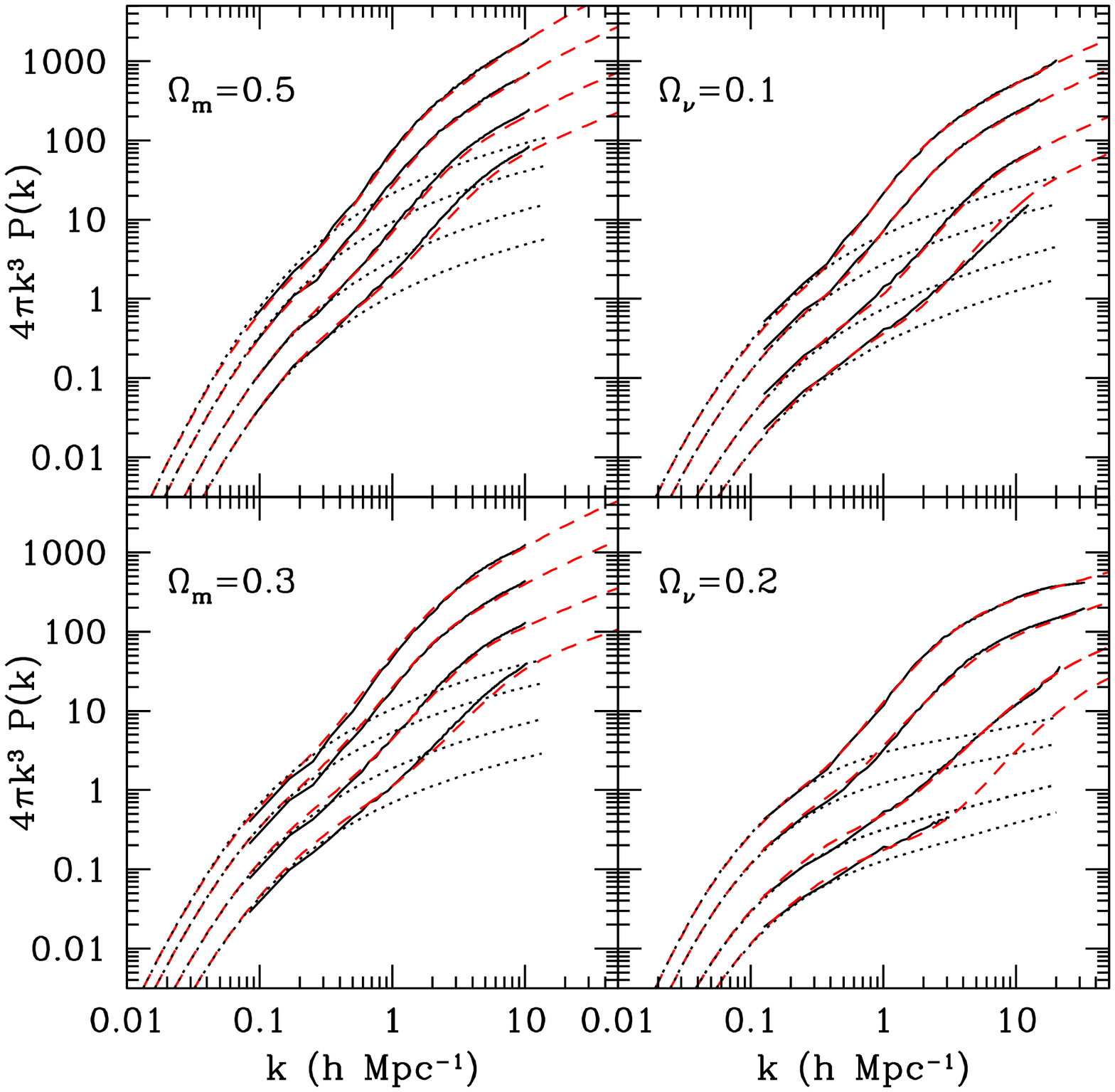}
\caption{The linear and fully evolved power spectrum at various
redshifts for two flat LCDM models ($\om=0.3$ and 0.5) and two flat
C+HDM models ($\onu=0.1$ and 0.2).  The solid curves are computed
directly from the simulations; the dashed curves show the close
approximation given by eq.~(\ref{master}) of this paper; the dotted
curves represent the linear power spectrum given by eqs.~(1) and (2).
All are normalized to the 4-year COBE data, and the present values of
$\sigma_8$ are 1.29 ($\om=0.3$), 1.53 ($\om=0.5$), 0.9 ($\onu=0.1$),
and 0.81 ($\onu=0.2$).  In each panel, the curves from bottom up are
for scale factors $a=0.2$, 0.33, 0.6, and 1.}
\end{figure}

\begin{figure}
\epsfxsize=6.5truein 
\epsfbox{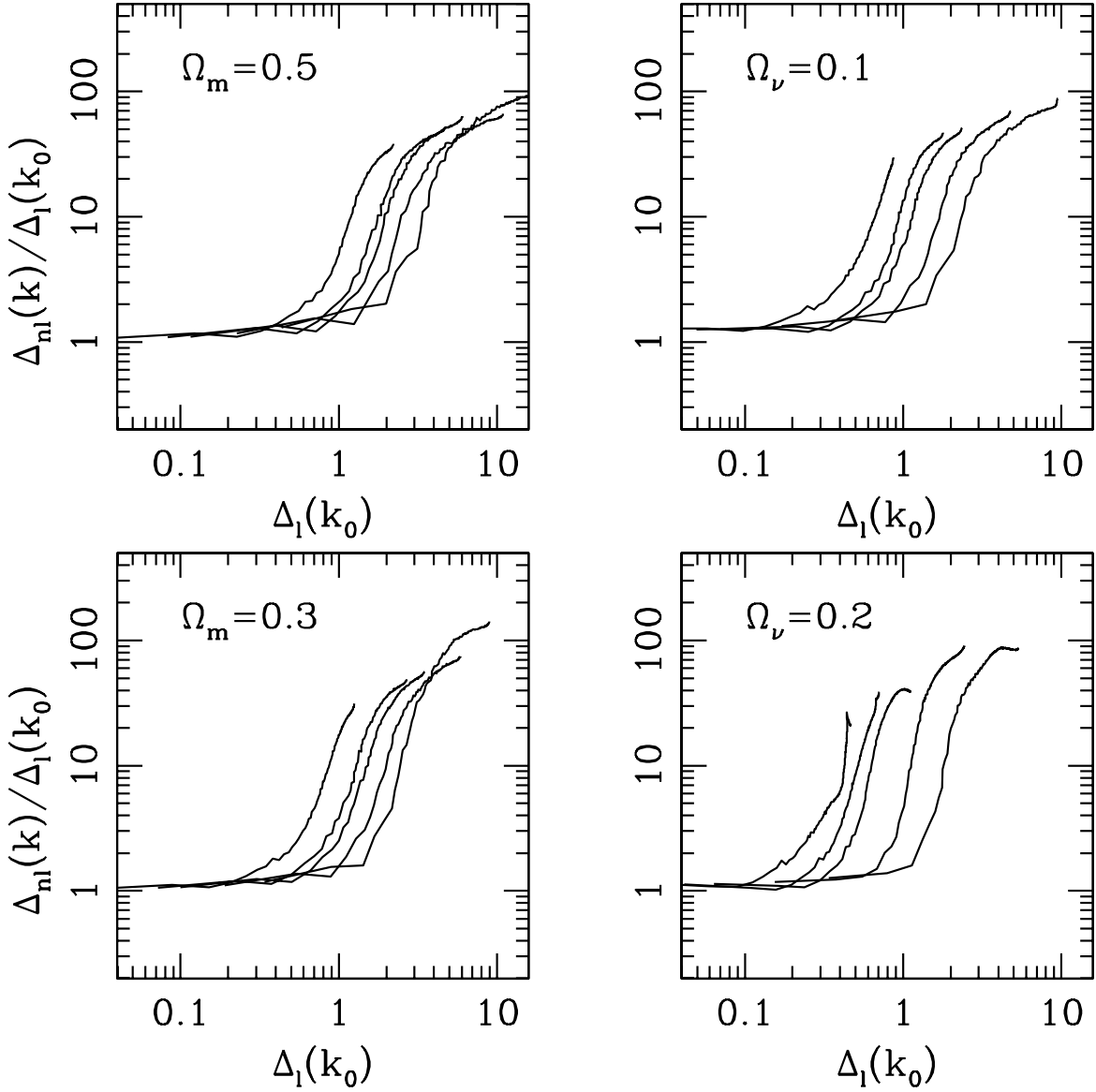}
\caption{Ratio of the nonlinear to linear density variance,
$\Dnl(k)/\Dl(k_0)$, as a function of $\Dl(k_0)$ at various redshifts
for two LCDM and two C+HDM models.  The wavenumber $k_0$ corresponds
to the pre-collapsed linear value of $k$, where
$k_0=k\,(1+\Dnl)^{-1/3}$.  In each panel, the curves from left to right
correspond to $a=0.2$, 0.33, 0.4, 0.6, and 1.}
\end{figure}

\begin{figure}
\epsfxsize=6.5truein 
\epsfbox{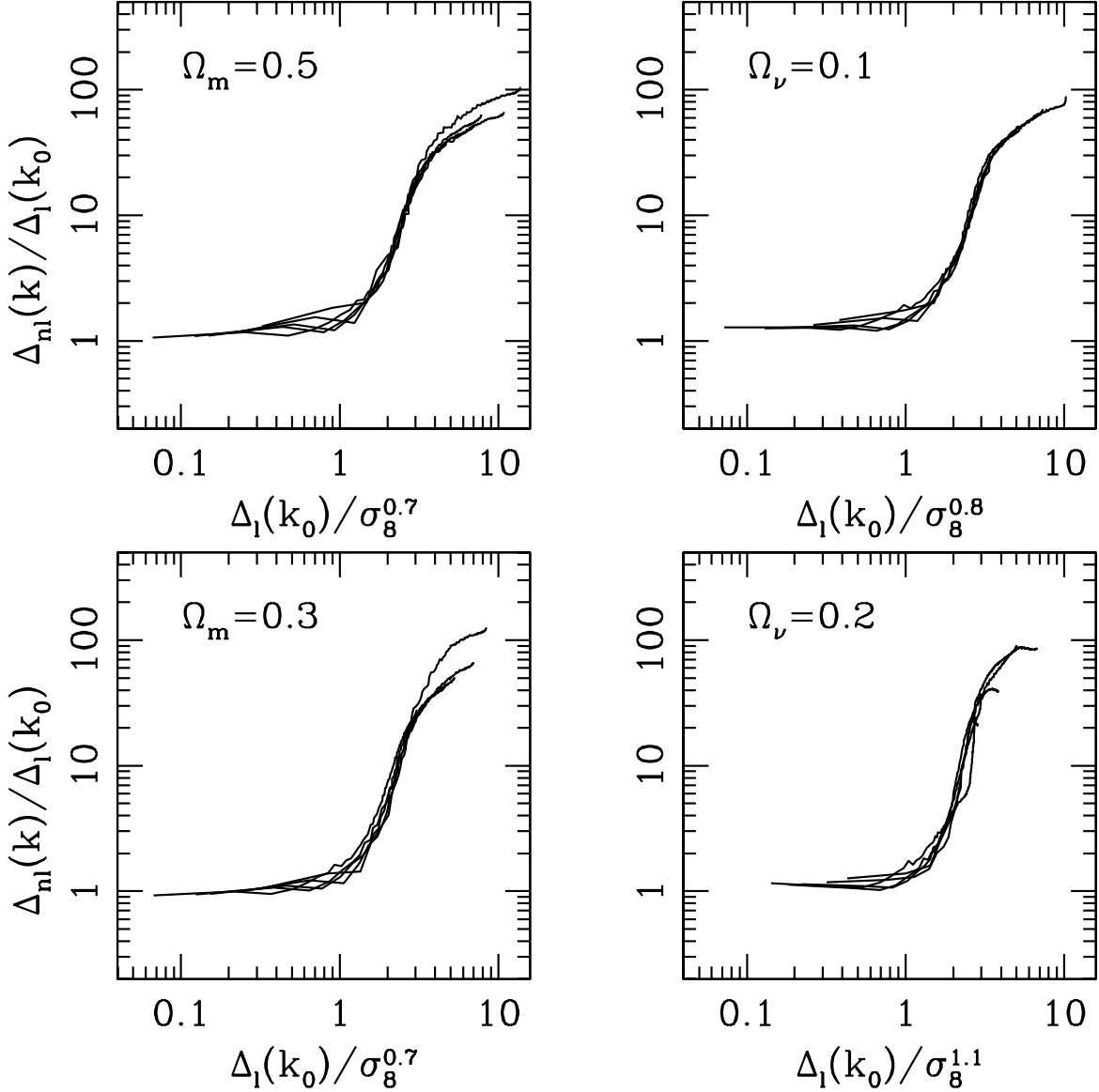}
\caption{Same as Figure 2, but the horizontal axis represents the
scaled $\Dl/\sigma_8^\beta$, where $\beta=0.7+10\,\onu^2$.  It shows
that the time dependence in the ratio $\Dnl/\Dl$ in Figure~2 is now
largely removed except at late times in the two LCDM models (see
text).}
\end{figure}

\begin{figure}
\epsfxsize=6.5truein 
\epsfbox{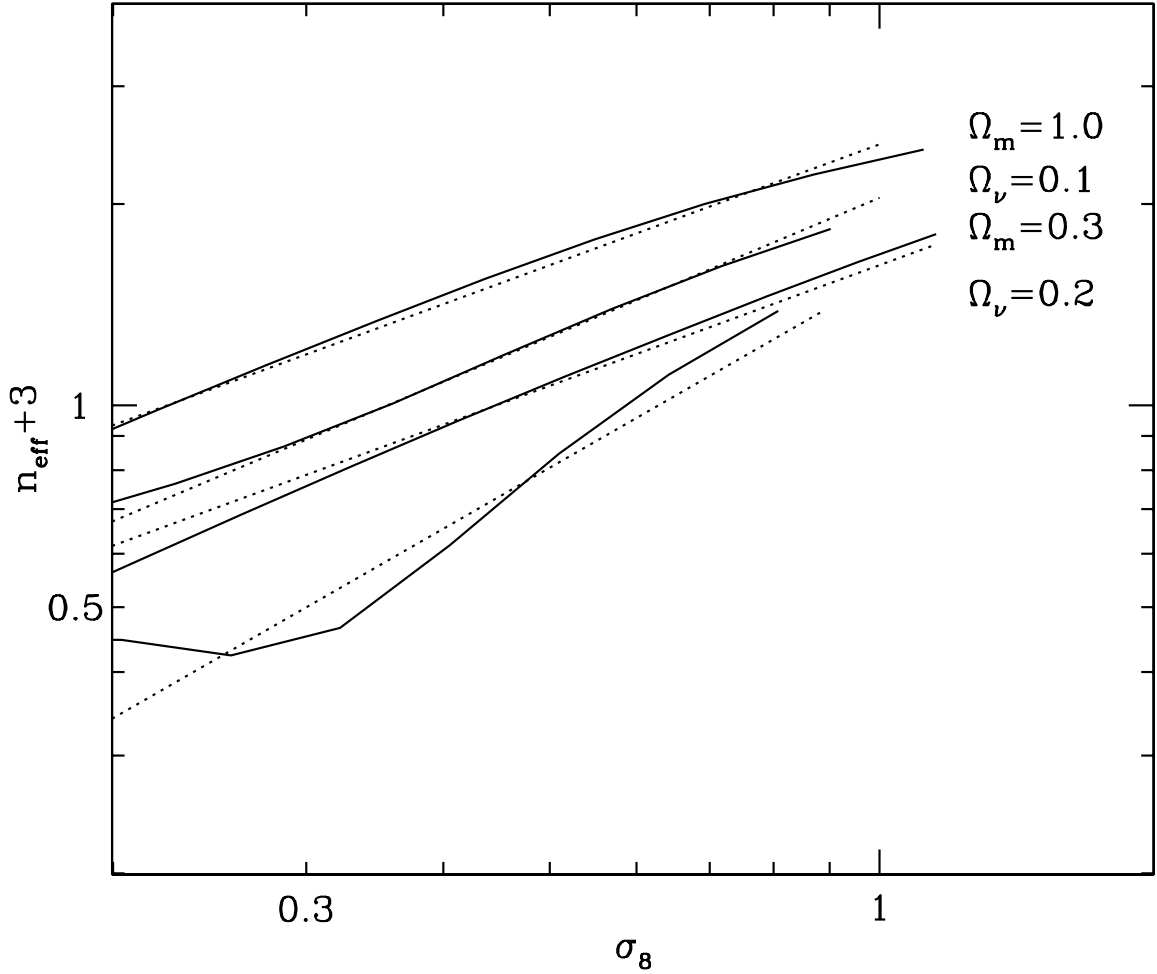}
\caption{The effective spectral index, $n_{\rm eff}$, as a function of
$\sigma_8$ for four models.  The dotted lines illustrate that the
power law $d\ln (n_{\rm eff}+3)/d\ln \sigma_8 \propto \beta$ is a good
approximation for $\sigma_8\go 0.3$.}
\end{figure}

\end{document}